\documentclass[12pt]{article}
\usepackage[margin=1in, paperwidth=8.5in, paperheight=11in]{geometry}
\usepackage{amsmath}
\usepackage{graphicx}%
\usepackage{amsfonts}%
\usepackage{amssymb}
\usepackage{color}
\usepackage{float,subfigure}
\usepackage[round]{natbib}
\usepackage{cite}
\usepackage{setspace}
\usepackage{comment}
\usepackage{url}

\newcommand{\bbeta}{ \mbox{\boldmath $ \beta $} }

\newcommand{\eps}{ \mbox{$\epsilon$}}

\newcommand{\btheta}{ \mbox{\boldmath $ \theta $} }

\newcommand{\bmu}{ \mbox{\boldmath $\mu$} }

\newcommand{\Sig}{ \mbox{$\Sigma$} }
\newcommand{\sig}{ \ensuremath{\sigma}}

\newcommand{\bzero}{\textbf{0}}

\newcommand{\bK}{ {\bf K} }

\newcommand{\bs}{ {\bf s} }

\newcommand{\bw}{ {\bf w} }

\newcommand{\bx}{ {\bf x} }

\newcommand{\bY}{ {\bf Y} }

\newcommand{\Cov}{\mbox{Cov}}
\newcommand{\Cor}{\mbox{Cor}}

\newcommand{\given}{\,\vert\,}

\newcommand{\Sigy}{\Sig_Y}

\newcommand{\Sigw}{\Sig_W}

\newcommand{\Sigya}{\Sig_{Y,(N)}}
\newcommand{\Sigwa}{\Sig_{W,(N)}}

\begin{document}

\thispagestyle{empty}
\setcounter{page}{0}

\begin{center}
{\Large \textbf{{Generating Partially Synthetic {Geocoded} Public Use Data with Decreased Disclosure Risk Using Differential Smoothing}}}

\bigskip

\textbf{Harrison Quick$^{1*}$, Scott H. Holan$^{2}$, Christopher K. Wikle$^{2}$}\\
$^{1}$ Division of Heart Disease and Stroke Prevention, Centers for Disease Control and Prevention, Atlanta, GA 30329\\
$^2$ Department of Statistics, University of Missouri, Columbia, Missouri.\\
$^{*}$ \emph{email:} HQuick@cdc.gov\\
\end{center}

{\textsc{Summary.} When collecting geocoded confidential data with the intent to disseminate, agencies often resort to altering the geographies prior to making data publicly available due to data privacy obligations. An alternative to releasing aggregated and/or perturbed data is to release multiply-imputed synthetic data, where sensitive values are replaced with draws from statistical models designed to capture important distributional features in the collected data. One issue that has received relatively little attention, however, is how to handle spatially outlying observations in the collected data, as common spatial models often have a tendency to overfit these observations. The goal of this work is to bring this issue to the forefront and propose a solution, which we refer to as ``differential smoothing.'' After implementing our method on simulated data, highlighting the effectiveness of our approach under various scenarios, we illustrate the framework using data consisting of sale prices of homes in San Francisco.
}


\newpage

\section{Introduction}\label{sec:intro}
When collecting confidential data with the intent to disseminate, there is often both an ethical as well as legal obligation for agencies to protect the privacy of data subjects' identities and sensitive attributes.
This charge can be particularly challenging for agencies who seek to include fine levels of geography (e.g., {latitude/longitude}) in the public use files they provide. While data users can benefit greatly from this detailed spatial information, this can also enable ill-intentioned users to identify individuals in the dataset. This disclosure risk can be especially high in 
regions where individuals with sensitive attributes may be more unique.

As a result, agencies often resort to altering (or worse, suppressing) the geographies and/or sensitive attributes before making data publicly available.  A common technique is to aggregate data from the individual level to areal units (e.g., Census tracts or counties).  Not only can this destroy the ability to estimate the spatial structure at finer geographies than the aggregate level, but it may also lead researchers to make ecological fallacies {\citep{freedman, lawsonetal,bradley2015regionalization}}. Agencies may also randomly move each record's observed location to another location, e.g., within some radius $r$ 
of the true location.  
In addition to having a negative impact on the spatial structure in the released data \citep[e.g.,][]{rushton,guttman}, the effect of this perturbation may be overlooked by researchers, potentially resulting in false conclusions.

An alternative to releasing aggregated and/or perturbed data is to release multiply-imputed synthetic data, where sensitive values are replaced with draws from statistical models designed to capture important distributional features in the collected data.  In some cases, agencies
may generate \emph{fully} synthetic data \citep{rubin93, reiter:2002, reiter:2002a,raghu:rubin:2001,QHWR}, in which the released datasets are comprised entirely of simulated records.  We, however, 
take a \emph{partially} synthetic approach in which only a collection of values/variables are replaced with imputed values 
\citep{little93,kennickel:1997, 
abowdwood04,reiter2003, reitermi,an:little:2007,toth:2014}.  {Specifically, we assume the data consist of exact geographic locations and covariate information for each individual, as well as a continuously varying response 
which will be multiply imputed.}

One issue that has yet to be adequately addressed, however, is how to handle spatially outlying observations in the collected data.
For instance, suppose the agency would like to release annual income data for individuals from a number of subpopulations for a given city.  Further, suppose a Census tract contains only one black female over 50 years of age.  Were the agency to release aggregate data, it is likely that this Census tract's income information would be suppressed for this particular subpopulation in order to protect this individual's privacy.  When generating (fully or partially) synthetic data, however, such steps to protect the individual's privacy may not even be {considered}, much less taken.  Furthermore, such a crude method is ignorant to the \emph{size} of a given areal unit --- e.g., the sole individual in an urban Census tract (where tracts may be more densely clustered) may in fact be at \emph{less} risk of disclosure than one of a handful of individuals in a rural Census tract which stretches over an area of several miles. As this issue is better illustrated in a partially synthetic framework, we focus here on the partially synthetic (henceforth referred to as simply ``synthetic'') case.
{That said, this issue still pertains to methods for generating fully synthetic data like \citet{QHWR}, though the impact is lessened due to the possibility of no synthetic observations near the locations of the spatial outliers.}

{We would be remiss not to mention the ``robust kriging'' literature, a concept proposed by {\citet{hawkins:cressie}}.  As discussed further by {\citet{nirel} and \citet{mugglestone}}, the goal of robust kriging is to
develop methods of obtaining parameter estimates which are robust to observations whose responses are outlying (or otherwise not in line with model assumptions).
This is in contrast to our focus here, where
we are concerned with observations whose \emph{locations} are considered outlying and how this relates to disclosure risk.}

The goal of this work is to bring this issue to the forefront and propose a solution.
We begin in Section~2 by illustrating, in detail, the potential risks and how existing approaches fail to address the root of the problem.
In Section~3, we extend existing methods for generating synthetic data to further reduce disclosure risk for spatially outlying observations using a concept we refer to as \emph{differential smoothing}.
We implement these methods on simulated data in Section~4, highlighting the effectiveness {of our approach} under various scenarios.  {We then apply the methodology to data consisting of sale prices of homes in San Francisco in Section~5.  While privacy is not necessarily an issue for these data, they serve as a reasonable surrogate for household-level data, where disclosure risks would be of chief concern.}  Finally, in Section~6, we provide concluding remarks and some ideas for {future research}.

\section{Potential Disclosure Risks in Synthetic Data}\label{sec:potential}
Before discussing the potential risks when generating synthetic data, we must first select a method for modeling the true data.  For the sake of illustration, we shall assume that the data consists of continuous responses (e.g., annual income) from a single population.  While datasets generally consist of data collected from multiple populations (e.g., race, socioeconomic status, etc.),
we will restrict our attention to the univariate case; the topic of joint modeling is discussed further in Section~\ref{sec:disc}.

Let $\bs_i$ and $Y(\bs_i)$ be the location and response variable for the $i$-th individual, for $i=1,\ldots,N$.  For a continuously varying $Y(\bs)$ and vector of model parameters, $\btheta$, {we may choose a model of the form}
\begin{equation}
Y(\bs_i){\vert\btheta} \sim N(\bx(\bs_i)'\bbeta + w(\bs_i),\tau^2) \label{eq:mody}
\end{equation}
where $\bx(\bs_i)$ is a vector of spatially varying covariates with a corresponding vector of regression coefficients, $\bbeta$, and $w(\bs_i)$ is a random effect that induces correlation between the responses.  To account for spatial correlation in the responses, {a highly} flexible option is to assume $w(\bs)$ is a mean-zero Gaussian process, $GP(0,K(\cdot,\cdot;\sig^2,\phi))$, where $K(\bs_i,\bs_j;\sig^2,\phi) = \Cov(w(\bs_i),w(\bs_j))$.  For a collection of spatial locations, ${\cal S} = \{\bs_1,\ldots,\bs_{N}\}$, we define $\bw = \{w(\bs_1),\ldots,w(\bs_{N})\}'$ and assume
$\bw \given \sig^2,\phi \sim {MVN}\left(\bzero,\Sig_W\left(\sig^2,\phi\right)\right)$,
where the $(i,j)$-th element of $\Sig_W\left(\sig^2,\phi\right)$ is $K(\bs_i,\bs_j;\sig^2,\phi)$. For the sake of brevity, we suppress the conditioning and simply write $K(\bs_i,\bs_j)$ and $\Sig_W$.
We define $\bK_i$ to be the $({N}-1)$-dimensional vector with components $K(\bs_i,\bs_j)$ for $i\ne j$. While there are numerous choices for $K(\cdot,\cdot)$, we will illustrate our approach using an exponential covariance structure where $\Cov(w(\bs_i),w(\bs_j)) = \sig^2 \exp\left\{-\phi||\bs_i-\bs_j||\right\}$.  Here, $\sig^2$ represents the variance of the spatial process and $\phi$ denotes the spatial range, yielding $\btheta=\left(\bbeta, \bw,\tau^2,\sig^2,\phi\right)$ as the parameters to be estimated.  When ${N}$ is large, inverting $\Sig_W$ can be computationally burdensome, and a low-rank approximation such as the modified predictive process \citep{banerjee2010} may be required.  {While the approach we describe in Section~\ref{sec:methods} can be implemented using a low-rank approximation,
for the sake of illustration,
we will assume $N$ is of a manageable size.  This will allow us to focus on the properties of our approach rather than details
of the low-rank approximation.
}

\begin{figure}[t] 
   \begin{center}
        \includegraphics[width=.05\textwidth]{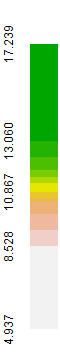}
        \subfigure[True Response]{\includegraphics[width=.3\textwidth]{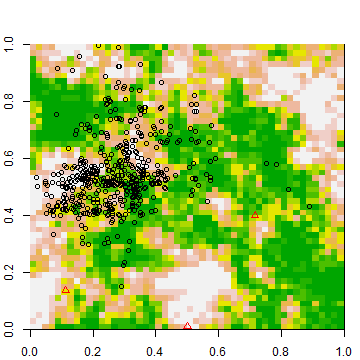}\label{fig:simtrue}}
        \subfigure[Estimated Response]{\includegraphics[width=.3\textwidth]{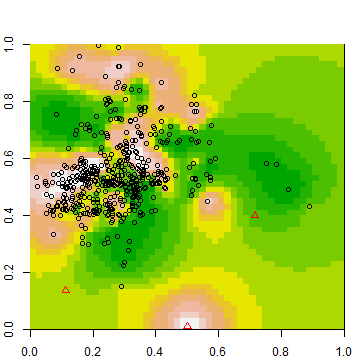}\label{fig:sim}}
        \subfigure[{Synthetic Responses} at $(0.51,0.01)$]{\includegraphics[width=.3\textwidth]{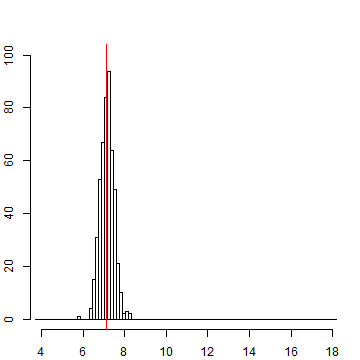}\label{fig:syn}}
    \end{center}
   \caption{Panels~(a) and (b) display the true and ({unrestricted}) estimated response surfaces for the data.  Locations are denoted by circles for non-at-risk individuals and red triangles for the at-risk individuals.  Panel~(c) displays the distribution of $L=500$ synthetic individuals at location $(0.51,0.01)$, generated using the surface in panel~(b).}
    \label{fig:potential}
\end{figure}

To illustrate the potential disclosure risk in synthetic data, we generate ${N}=500$ observations from~(\ref{eq:mody}) where $\tau^2 = 0.0625$, $\sig^2 = 4$, $\phi=12.7$, and locations on the unit square; the individuals are shown in Figure~\ref{fig:simtrue}, overlaid on the true response surface.  This choice for $\phi$ corresponds to {$\Cor(w(\bs_i),w(\bs_j)) < 0.05$ for $||\bs_i-\bs_j||>\sqrt{2}\slash 6 \approx 0.23$}, {and the values for $\tau^2$ and $\sig^2$ were chosen such that the ratio of $\sig^2$ to $\tau^2$ was large---the impact of this ratio can be seen in Section~\ref{sec:ai}}.  The observation at location $(0.51,0.01)$ in {Figure~\ref{fig:simtrue}} 
is further than 0.26 units away from the remaining 499 observations, and is henceforth referred to as the ``spatial outlier.'' Without loss of generality,
we 
assume this is the ${N}$-th observation in the {dataset. Later in Section~\ref{sec:ai}, we will identify spatial outliers using a more relaxed definition.}

To model these data, we may use an intercept-only model, assume an exponential covariance structure for the spatial random effects, and take a Bayesian approach, completing the model specification by defining vague priors for the model parameters.
After fitting the Bayesian hierarchical model and obtaining posterior distributions for the parameters, we achieve the estimated response surface shown in Figure~\ref{fig:sim}.  Of particular importance here is the presence of a ring encircling the spatial outlier, around which the predicted values appear to gradually decrease from the estimate of $\widehat{\beta}_0=11.44$ outside the ring to $\widehat{Y}(\bs_N) = \widehat{\beta}_0 + \widehat{w}(\bs_N)=7.16$, which may be considered too close to the true value of 7.08.

Given our existing spatial locations, we can generate $L=500$ partially synthetic datasets by sampling synthetic responses, denoted $Y(\bs_i)^{\dagger(\ell)}$, from the posterior predictive distribution
\begin{equation*}
Y(\bs_i)^{\dagger(\ell)} \given \btheta^{(\ell)} \sim N\left(\beta_0^{(\ell)} + w(\bs_i)^{(\ell)}, \left\{\tau^{(\ell)}\right\}^2\right)
\end{equation*}
using the {methods described in \citet{QHWR} for marked point processes},
where $\btheta^{(\ell)}$, $\beta_0^{(\ell)}$, $w(\bs_i)^{(\ell)}$, and $\tau^{(\ell)}$ denote the $\ell$-th approximately independent samples from the respective posterior distributions for $\ell=1,\ldots,L$, and $i=1,\ldots,N$.  Figure~\ref{fig:syn} displays a histogram of the 500 synthetic responses for the spatial outlier.  Alarmingly, this empirical distribution is almost perfectly centered around the true value for $Y(\bs_N)=7.08$, denoted by the red line.

In essence, fitting a spatial model for data with spatial outliers may lead to overfitting in the vicinity of the outliers.  {Furthermore, non-model-based methods for smoothing 
may also yield potentially unsatisfactory results. 
For instance,
\citet{ZhouEtAl2010} show that replacing $Y(\bs_i)$ with 
$\widetilde{Y}(\bs_i) = \sum_{k=1}^N  W(\bs_i,\bs_k) Y(\bs_k)$ --- where $W(\cdot,\cdot)\ge0$ is some spatially-associated weight function with $\sum_{k=1}^N W(\bs_i,\bs_k) =1$ --- can produce synthetic data with decreased risk.
Unfortunately, if $\bs_N$ 
is a spatial outlier, this can still result in $\widetilde{Y}(\bs_N) \approx Y(\bs_N)$ when $W(\bs_i,\bs_k)\approx0$ for $k\ne i$ for a {distance-based} choice of $W(\cdot,\cdot)$.
While this could be avoided 
by imposing a ``disclosure constraint,''
this may be
detrimental to the remaining
observations.
Needless to say, this is a problem that is easy to overlook yet difficult to fully address.
}

\section{Differential Smoothing Framework}\label{sec:methods}
\subsection{Background for Bayesian spatial models}
Using the model in~(\ref{eq:mody}), we can write
$\bY\given\btheta \sim N(\bmu+\bw,\Sig_Y)$
and $\bw\given\sig^2,\phi \sim N(\bzero,\Sig_W)$ 
where $\bY = \{Y(\bs_1),\ldots,Y(\bs_N)\}'$, $\bmu=\{\mu(\bs_1),\ldots,\mu(\bs_N)\}'$, $\mu(\bs_i)=\bx(\bs_i)'\bbeta$, and $\Sig_Y$ is a diagonal matrix with elements $\tau^2$.  {We can then show that the full conditional {distribution} for $\bw$ is} 
\begin{equation}\label{eq:full}
\bw\given\cdot \sim N\left(\left[\Sigy^{-1}+\Sigw^{-1}\right]^{-1}\Sigy^{-1}(\bY-\bmu),\left[\Sigy^{-1}+\Sigw^{-1}\right]^{-1}\right).
\end{equation}
To fit this model under a Bayesian framework, we must specify prior distributions for our remaining model parameters: $\bbeta$, $\sig^2$, $\phi$, and $\tau^2$.

Again, we suppose that the $N$-th observation is a spatial outlier---and thus is determined to have a high disclosure risk---while the remaining $N-1$ observations are clustered together and treated as having no disclosure risk.  In order to account for this in the model, we define ``risk weights'' $a_i\in[0,1]$ which will be used to differentially smooth the predicted surfaces.  We then define the diagonal matrix $A$ as having elements
$A_{ii} = 1\slash{\sqrt{1+\gamma a_i}}$
where $\gamma\ge 0$ denotes a ``global risk'' parameter, and define $\bw^* = A\bw$. 
Now, if we want to partition {the observations} based on risk, we would have
\begin{equation} \label{eq:part}
\bw^* \given \sig^2,\phi
\sim N\left(
\begin{pmatrix}
\bzero\\
0
\end{pmatrix},
\begin{bmatrix}
A_{(N)}\Sigwa A_{(N)} & A_{(N)}\bK_N\slash\sqrt{1+\gamma a_N}\\
\bK_N' A_{(N)}\slash\sqrt{1+\gamma a_N} & \sig^2\slash(1+\gamma a_N)
\end{bmatrix}\right),
\end{equation}
where $A_{(N)}$ and $\Sigwa$ denote the $(N-1)\times(N-1)$ matrices constructed by removing the last row and column of $A$ and $\Sigw$, respectively.

\subsection{Defining the $a_i$ and $\gamma$}\label{sec:ai}
{
Rather than define $a_i$ on a continuum, a simple option is to let
$a_i=1$ if the $i$-th observation is deemed a spatial outlier and $a_i=0$ otherwise.  For instance, we may consider the $i$-th observation as an outlier if the distance to the nearest neighbor, $\min_{j\ne i}||\bs_i-\bs_j|| \ge M$ for some $M$.  To define $M$, we may choose a specification based on an inversion of the correlation structure used, such as $M(\phi) \ge -(\log 0.20)\slash \phi$ --- which ensures that $\max_{j\ne i}\Cor\left(w(\bs_i),w(\bs_j)\right) \ge 0.20$ for non-outliers.  While there is no theoretical basis for this choice, we have found that it offers a compromise between the utility and the disclosure risk of the synthetic data we generate.
Updating~(\ref{eq:part}) with this restriction yields
\begin{equation} \label{eq:parta}
\bw^* \given \sig^2,\phi
\sim N\left(
\begin{pmatrix}
\bzero\\
0
\end{pmatrix},
\begin{bmatrix}
\Sigwa & \bK_N\slash\sqrt{1+\gamma}\\
\bK_N' \slash\sqrt{1+\gamma} & \sig^2\slash(1+\gamma)
\end{bmatrix}\right).
\end{equation}
We discuss the topic of continuous-valued $a_i$ later in Section~\ref{sec:disc}.

Choosing a value for $\gamma$ can be less clear, so first we need to investigate how different values for $\gamma$ affect the model.
To better elucidate this, suppose $\bs_N$ is sufficiently far away from the other points such that $\exp(-\phi||\bs_N - \bs_j||)\approx0$ for $j\ne N$; i.e.,
\begin{equation*}
\Cov\left(\bw_{(N)}^*,w(\bs_N)^*\right) =
\begin{bmatrix}
\Sigwa & \bzero\\
\bzero' & \sig^2\slash(1+\gamma)
\end{bmatrix}.
\end{equation*}
Plugging this into the full conditional distribution for $\bw^*$ (which takes the form of~(\ref{eq:full}) with $\Sigw$ replaced by $A\Sigw A$) yields
\begin{align*}
E[\bw^*\given\cdot] &=
\begin{bmatrix}
\left[\Sigya^{-1} + \Sigwa^{-1}\right]^{-1}\Sigya^{-1} & \bzero\\
\bzero' & \frac{\sig^2\slash(1+\gamma)}{\tau^2 + \sig^2\slash(1+\gamma)}
\end{bmatrix}
\begin{pmatrix}
\bY_{(N)} - \bmu_{(N)}\\
Y(\bs_N) - \mu(\bs_N)
\end{pmatrix}
\\
\text{and}\;\;\;
V[\bw^*\given\cdot] &=
\begin{bmatrix}
\left[\Sigya^{-1} + \Sigwa^{-1}\right]^{-1}\Sigya^{-1} & \bzero\\
\bzero' & \frac{\sig^2\slash(1+\gamma)}{\tau^2 + \sig^2\slash(1+\gamma)}
\end{bmatrix}.
\end{align*}
Note that this implies that the conditional expected value of $w^*(\bs_N)$ is a weighted average of $Y(\bs_N) - \mu(\bs_N)$ and the prior mean of 0; i.e.,
\begin{align}
E\left[w^*(\bs_N)\given\cdot\right] &= \frac{\sig^2\slash(1+\gamma)}{\tau^2 + \sig^2\slash(1+\gamma)} \left(Y(\bs_N)-\mu(\bs_N)\right) + \frac{\tau^2}{\tau^2 + \sig^2\slash(1+\gamma)} (0)\notag\\
&= \alpha \left(Y(\bs_N)-\mu(\bs_N)\right) + (1-\alpha) (0),\label{eq:alpha}
\end{align}
where $\alpha\in \left[0,\sig^2\slash(\sig^2+\tau^2)\right]$ denotes the {degree of spatial smoothing}.
Note that setting $\gamma=0$ yields $\alpha = \sig^2\slash(\sig^2+\tau^2)$, which results in the standard unrestricted model.
When choosing a non-zero, finite value for $\gamma$, one option may be to force $\alpha$ to take some value in $\left(0,\sig^2\slash[\sig^2+\tau^2]\right)$ to achieve a desired level of differential smoothing. For instance, if $\alpha=1\slash2$, this corresponds to $\gamma = \sig^2\slash\tau^2 -1$, provided $\sig^2 > \tau^2$.
To achieve a ``fully smoothed'' process for our outlying observations, however, we let $\alpha = 0$, which corresponds to $\gamma=\infty$.  Furthermore, note that this restriction forces $E[w^*(\bs_N)\given\cdot]= V[w^*(\bs_N)\given\cdot]= 0$; i.e., if we let $\gamma = \infty$, this implies $w^*(\bs_N)\equiv 0$
(note that $w^*(\bs_N)\equiv 0$ does not imply $w(\bs_N) \equiv 0$). 

}

\subsection{Implementation}
To implement our differential smoothing approach, we first fit the {unrestricted} model:
\begin{align}\label{eq:hier}
\pi(\bbeta, \bw, \sig^2, \phi, \tau^2 \given \bY) \propto& N(\bY\given \bmu +\bw,\Sigy)  \times N(\bw\given \bzero,\Sigw) \times \pi(\bbeta,\sig^2,\phi,\tau^2),
\end{align}
with $a_i=0$ for all $i$ (or $\gamma=0$) and using vague prior specifications for $\bbeta$, $\sig^2$, $\phi$, and $\tau^2$, where $\pi(x\given y)$ denotes the conditional distribution of $x$ given $y$.  
We could then specify $a_i$ and $\gamma$ to remain functions of our model parameters (i.e., $a_i(\phi)$ and $\gamma(\sig^2,\tau^2)$), changing the degree of smoothing adaptively.  As we will discuss in Section~\ref{sec:disc}, however, this may have consequences regarding parameter estimation (e.g., the loss of conjugacy for $\sig^2$), 
and thus we do not pursue this here.  Instead, we specify the $a_i$ using the distance to the nearest neighbor (as a function of the posterior median of $\phi$ from our unrestricted model)
and implement a fully smoothed restriction by setting $\gamma=\infty$. We then fit the restricted hierarchical model
\begin{align}\label{eq:restrict}
\pi(\bbeta, \bw, \sig^2, \phi, \tau^2 \given \bY) \propto& N(\bY\given \bmu +A\bw,\Sigy)  \times N(\bw\given \bzero,\Sigw) \times \pi(\bbeta,\sig^2,\phi,\tau^2),
\end{align}
using these values of $a_i$ and $\gamma$.  To facilitate faster convergence, we can use samples from the unrestricted model as initial values for the restricted model, and we recommend fixing $\phi$ so as not to affect which observations are to be deemed ``spatial outliers''.


Using the samples drawn from the posterior distribution based from the restricted model, we then generate synthetic data from $Y(\bs)^{\dagger(\ell)} \given \bmu^{(\ell)}, w(\bs)^{(\ell)}, \tau^{(\ell)} \sim N\left(\bmu^{(\ell)} + w(\bs)^{(\ell)}, \left\{\tau^{(\ell)}\right\}^2\right)$.
Here again, note that if we use the fully smoothed approach where $\gamma=\infty$, the $w(\bs_N)^{(\ell)}$ are simply draws from the conditional prior distribution, $w(\bs_N)\given \bw_{(N)}$.

\section{Simulated Example}\label{sec:ex}
Before delving into an assessment of the proposed method, we will first describe the motivation for the simulated example used both here and in Section~\ref{sec:potential}.  The response is intended to correspond to an individual's log-transformed income, centered around an annual income of roughly \$50,000 with a small proportion of the sample having incomes higher than \$1,000,000 and some individuals having incomes below the poverty line.  The observations are sampled such that the majority of the data come from a high density region of the spatial domain, while a few of the individuals reside in less densely populated regions (with respect to the subpopulation being sampled).  {As is common with real data}, the simulated data contain pockets of both high and low income individuals 
{(in practice, agencies tend to release top-coded income data \citep[e.g., see][]{topcode}, another data-privacy method which can result in bias)}. 
{To achieve this in these}
data, we generated from the model
where 
$\tau^2=0.0625$, $\bw\given\sig^2,\phi \sim N(0,\Sigw)$ with $\sig^2=4$ and $\phi=12.7$, and
\begin{equation}
Y(\bs_i)\given w(\bs_i),\tau^2 \sim N(11 + 0.25 \times ||s_{i1}-0.25|| + 0.25 \times ||s_{i2}-0.5|| + w(\bs_i),\tau^2),
\end{equation}
As displayed in Figure~\ref{fig:simtrue}, we observe a spatially outlying individual at $(0.51,0.01)$ in a relatively low income bracket who we have identified as being at-risk for disclosure.  Using the methods described in Section~\ref{sec:methods}, we will demonstrate our differential smoothing approach for protecting this and other individuals.  {We will also compare these results to those from an analysis where the spatial outlier was removed from the data prior to model fitting.}

\begin{figure}[t] 
   \begin{center}
        \subfigure[Unrestricted]{\includegraphics[width=.3\textwidth]{sim_plot1.png}\label{fig:simplot1}}
        \subfigure[Restricted]{\includegraphics[width=.3\textwidth]{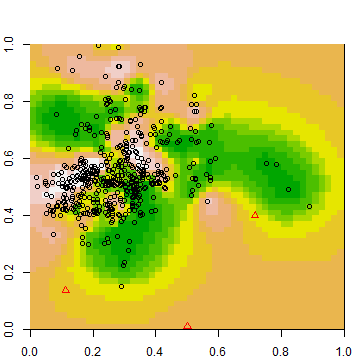}\label{fig:simplot3}}
        \includegraphics[width=.05\textwidth]{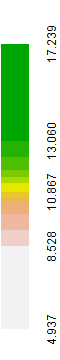}
    \end{center}
   \caption{Estimated response surfaces from the unrestricted and restricted models using the simulated data.}
    \label{fig:simsurface}
\end{figure}

After fitting the unrestricted hierarchical model in~(\ref{eq:hier}),
we consider the restricted model of Section~\ref{sec:methods}, where we let $a_i$ be a 0/1 indicator function for the absence of neighbors within {$M=-\log(0.20)\slash\phi = 0.13$ units}, {resulting in 2 additional at-risk individuals} (denoted using red triangles in Figure~\ref{fig:simsurface}).  We then let $\gamma=\infty$, {forcing $w^*(\bs_i)\equiv0$} for the at-risk observations, while leaving the non-at-risk observations relatively unaffected.  Refitting the model under this specification, we obtain the estimated response surface in Figure~\ref{fig:simplot3}.  Comparing this figure to the unrestricted surface in Figure~\ref{fig:simplot1}, a number of features are noticeable.  First, as shown in Table~\ref{tab:simres1}, the estimate of $\beta_0$ has decreased from $11.35$ 
to $10.31$, largely due to the negative pull of the outlying observation at $(0.51,0.01)$, resulting in lower predictions for the unobserved regions on the right side of the spatial domain. Secondly, the ring of low predicted values around the spatial outlier has vanished, resulting in a surface that is essentially naive to the existence of this individual.  {Additionally, note that the estimate of $\sig^2$ in our restricted model is similar to that from the analysis of the suppressed dataset, while the estimates of $\beta_0$ and $\tau^2$ differ substantially.  This is because we cannot learn about $w(\bs_N)$ in either model, leaving $\beta_0$ and $\epsilon(\bs_N)$ to do more work in the restricted model.}

\begin{table}[t]
\begin{center}
\small
\begin{tabular}{|l|c|c|c|}
\hline
Model & $\beta_0$ & $\sig^2$ & $\tau^2$\\
\hline
Full Unrestricted   & 11.35 (11.13, 11.71)	& 3.83 (3.22, 4.59)	& 0.06 (0.05, 0.08)\\
Restricted    & 10.31 (10.20, 10.77)	& 3.77 (3.07, 4.53)	& 0.12 (0.10, 0.15) \\
Suppressed          & 11.71 (11.49, 12.07)  & 3.81 (3.19, 4.58)	& 0.06 (0.05, 0.08)	\\
\hline
\end{tabular}
\end{center}
\caption{Parameter estimates from each of our hierarchical models.  Note the effect of the spatial outlier whose value (7.06) is much less than the mean of the data (10.76).}
\label{tab:simres1}
\end{table}

\begin{figure}[t] 
   \begin{center}
        \subfigure[Unrestricted]{\includegraphics[width=.45\textwidth]{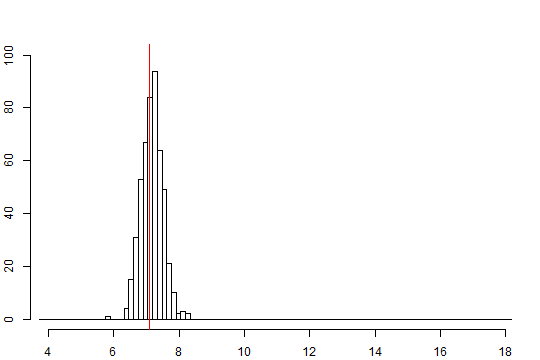}\label{fig:synplot1}}
        \subfigure[Restricted]{\includegraphics[width=.45\textwidth]{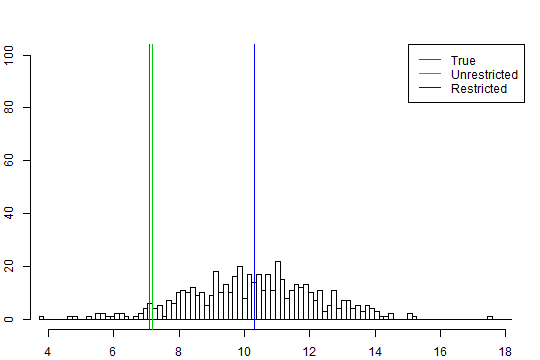}\label{fig:synplot3}}
    \end{center}
   \caption{Distributions of the synthetic responses for the spatial outlier from the unrestricted and restricted models using the simulated data.}
    \label{fig:simsyn}
\end{figure}

We now turn our attention to the synthetic data generated from these models.  Figure~\ref{fig:simsyn} displays the distributions of the synthetic responses for the spatial outlier.  In each panel, the true value for this individual is denoted by the red vertical line, while the histogram for the restricted model also contains a green line denoting the mean of the unrestricted synthetic responses (for comparison purposes) and a blue line denoting the mean for the set of restricted responses.  Here, we see the impact of the smoothing techniques in the restricted model, as now the synthetic responses are centered around the estimate for $\beta_0$ from Table~\ref{tab:simres1} instead of the true value of $7.08$.  Recalling that these responses are modeled after log-transformed annual incomes, we can assess the disclosure risk for this individual by computing the proportion of synthetic incomes within a certain $\epsilon$ of the truth \citep[see, e.g.,][]{QHWR}.  {For instance, 100\% of the synthetic incomes from the unrestricted model are within \$10,000 of the true value, compared to only 30\% for our restricted model.}  {Similarly, the proportion of synthetic incomes within 10\% of their true values for our three at-risk individuals has been reduced by at least 73\% and by an average of 20\% for the non-at-risk individuals.}
{To assess the utility of the synthetic data from our models, we fit
\begin{equation*}
Y^{\dagger(\ell)}(\bs_i) = \beta_0^{\dagger(\ell)} + \beta_1^{\dagger(\ell)}||s_{i1}-0.25|| + \beta_2^{\dagger(\ell)}||s_{i2}-0.5|| + \eps(\bs_i)^{\dagger(\ell)}
\end{equation*}
for $\ell=1,\ldots,L$ and used the combination rules in \citet{reiter2003} to obtain point and interval estimates for our regression parameters from each model.  Table~\ref{tab:simres2} displays these results for our unrestricted and restricted models, as well as those corresponding to the analysis of the suppressed data.  In general, our regression parameters, $\bbeta^{\dagger}$, are relatively unaffected, though this is not surprising given the small number of at-risk observations.}

\begin{table}[t]
\begin{center}
\small
\begin{tabular}{|l|c|c|c|}
\hline
Parameter & Full Unrestricted & Restricted & Suppressed Unrestricted\\
\hline
$\beta_0^{\dagger}$ (Intercept) & 10.58 (10.31, 10.85)	&	10.57 (10.3, 10.84)	&	10.54 (10.27, 10.80)	\\
$\beta_1^{\dagger}$ ($s_{i1}$ slope) & -0.16 (-0.33, 0.02)	&	-0.16 (-0.34, 0.02)	&	-0.14 (-0.32, 0.04)	\\
$\beta_2^{\dagger}$ ($s_{i2}$ slope) & 0.37 (0.20, 0.55)	&	0.40 (0.22, 0.57)	&	0.42 (0.24, 0.60)	\\
$Y^{\dagger}\left(\bs_N\right)$ & 7.18 (6.57, 7.80)	&	10.33 (6.21, 13.91)	&	11.86 (8.26, 15.63)	\\
\hline
\end{tabular}
\end{center}
\caption{Parameter estimates from the simulated example.
Note: the estimates for $\bbeta^{\dagger}$ from the unrestricted model mirror those from a fit of the real data, thus these results have not been shown for the sake of brevity.}
\label{tab:simres2}
\end{table}


\section{Real Data Example}\label{sec:sf}
Having illustrated the potential risks of the common, unrestricted model and demonstrating the effectiveness of our differential smoothing approach, we now look to apply our methodology to a dataset of home sale prices in San Francisco for the period from Feb.\ 2008 to July 2009.  These data were collected and described by \citet{adler} and consist of the sale price, the square footage, the number of bedrooms, and the spatial location (latitude and longitude) for each home.  For the purposes of this paper, we will restrict our attention to 
the 214 homes with one bedroom.  While these data themselves are not considered ``at-risk'' for disclosure (e.g., home listings are publicly available), the number of bedrooms and the home value may reasonably be considered as surrogates for sensitive household information such as the size of a household and the total household income, respectively. {Thus, we believe the dependencies
underlying these data are representative of those underlying data for which disclosure risk would be of concern.}

\begin{figure}[t] 
   \begin{center}
        \subfigure[Unrestricted]{\includegraphics[width=.3\textwidth]{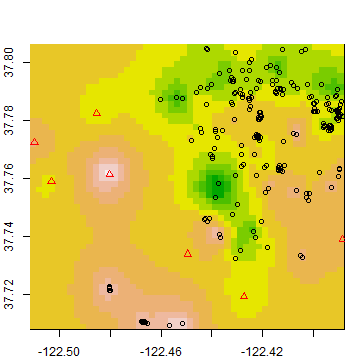}\label{fig:sfplot1}}
        \subfigure[Restricted]{\includegraphics[width=.3\textwidth]{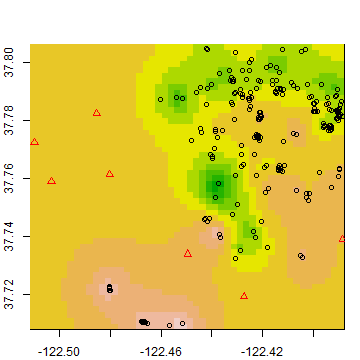}\label{fig:sfplot3}}
        \includegraphics[width=.05\textwidth]{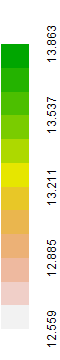}
    \end{center}
   \caption{Estimated response surfaces from the unrestricted and restricted models using the San Francisco home sales data.} 
    \label{fig:sfsurface}
\end{figure}

\begin{figure}[t] 
   \begin{center}
        \subfigure[Unrestricted]{\includegraphics[width=.45\textwidth]{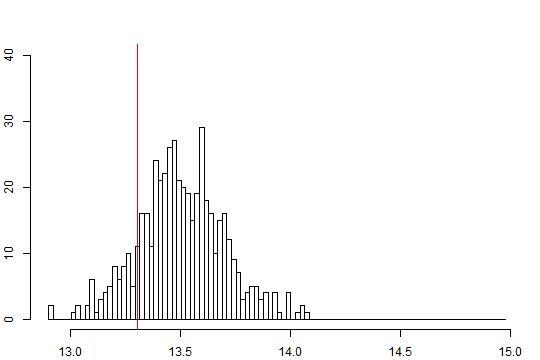}\label{fig:sfsynplot1}}
        \subfigure[Restricted]{\includegraphics[width=.45\textwidth]{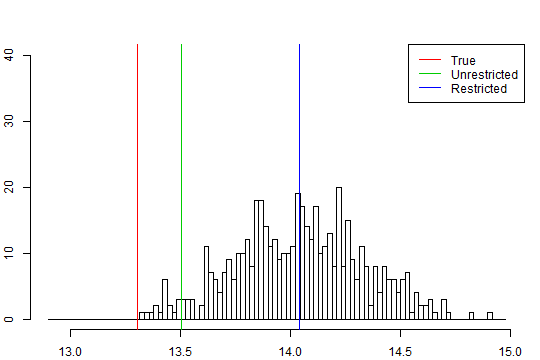}\label{fig:sfsynplot3}}
    \end{center}
   \caption{Distributions of the log-transformed synthetic sale prices for the home at $(-122.48, 37.76)$ from the unrestricted and restricted models using the San Francisco home sales data.}
    \label{fig:sfsyn}
\end{figure}

Following the process used in Section~\ref{sec:ex}, we first model the log-transformed sale prices using the unrestricted hierarchical model in~(\ref{eq:hier}) using the square footage as a covariate, yielding the prediction surface for $w(\cdot)$ in Figure~\ref{fig:sfplot1}.  Here again, we see ``rings'' in the prediction surface surrounding a number of potential spatial outliers (denoted by red triangles).  Based on the results presented in Section~\ref{sec:potential}, one can intuit that synthetic responses generated from this prediction surface for these outliers may be unacceptably close to their true values, thus motivating the use of differential smoothing.  Fortunately, the ratio of $\sig^2$ ($\approx 0.13$) to $\tau^2$ ($\approx 0.043$) is not as dramatic as in our simulated example, so the synthetic responses for the outlying observations are slightly shifted away from
$Y(\bs_i)=13.30$ toward $\bx(\bs_i)'\bbeta=14.15$, as shown in Figure~\ref{fig:sfsynplot1} for the observation at $(-122.48, 37.76)$.

We then proceed to fit the restricted model.  
Based on the distances to their nearest neighbors, {we identify seven homes as spatial outliers}.  Using this approach, we obtain the predicted surface for $w(\cdot)$ in Figure~\ref{fig:sfplot3} and the synthetic data in Figure~\ref{fig:sfsynplot3}.  As in the simulated example, this approach yields synthetic responses centered around the estimated value of $\bx(\bs_i)'\bbeta = 14.04$ in the restricted model.
To quantify this in terms of 
the risk of disclosure, the percentage of synthetic responses for the observation at $(-122.48, 37.76)$ which are within 10\% of the true value has been reduced by 93\% --- dropping from 46.2\% of our synthetic responses in the unrestricted model to just 3\% in our restricted model.  {Overall, this risk was reduced 50\% for at-risk individuals and 11\% for non-at-risk individuals.}

Now, in order for our restricted model to be a valuable tool, it is important to demonstrate that it can provide synthetic data which yield statistical inference similar to that from the real data.  To evaluate the utility of our synthetic data, we fit
\begin{equation*}
Y^{\dagger(\ell)}(\bs_i) = \beta_0^{\dagger(\ell)} + \beta_1^{\dagger(\ell)}\text{SqFt}(\bs_i) + \eps(\bs_i)^{\dagger(\ell)}
\end{equation*}
for $\ell=1,\ldots,L$ for each set of synthetic responses and again used the combination rules in \citet{reiter2003} to obtain point and interval estimates for our regression parameters.  Here, our results are even more impressive than in Table~\ref{tab:simres2}, as our restricted synthetic data produce estimates $\beta_0 =$ 13.233 (13.197, 13.269) and $\beta_1 =$ 0.269 (0.233, 0.306) --- estimates which are each within 0.002 of those from the real data.  To put these results in context, consider that the estimate for $\beta_1$ obtained from synthetic data generated from a model using a suppressed dataset is 0.279 (0.244, 0.314).

\section{Discussion}\label{sec:disc}
In this paper, we have shed light on a unique issue regarding disclosure risk encountered when generating spatially-referenced synthetic microdata from a population with spatially outlying observations.  After first illustrating an example of when this risk can arise in Section~\ref{sec:potential}, we proposed a framework which could be used to alleviate the risk of disclosure by restricting the hierarchical model using differential smoothing.  We then demonstrated its use on simulated data and applied it to a dataset of home sale prices in San Francisco.

Along with producing data which limit the risk of disclosure, producing data with high utility is of the utmost importance.  While the synthetic data that we have generated in Sections~\ref{sec:ex} and~\ref{sec:sf} have been able to provide
inference which was on par with those from the real data,
this is a much more nuanced problem in practice.  For instance, suppose our data consist of the gross annual household incomes for households in a particular region (and for the sake of illustration, suppose these data are \emph{not} top-coded).  If many of our spatial outliers also happen to be high earners (say, household incomes greater than \$250,000 per year), 
our synthetic data will likely \emph{underestimate} the number of high earners in the population. Fortunately, such issues can be addressed by constructing our {hierarchical models} based on important questions of inferential interest.  If we desire synthetic data which preserve the number of households in certain income brackets, we can specify \emph{conditional} models such as
\begin{equation}
Y(\bs_i) \given Y(\bs_i)\in G_k, \bbeta, \bw,\tau^2 \sim N\left(\bx(\bs_i)\bbeta + w(\bs_i),\tau^2\right) \times I\left\{Y(\bs_i) \in G_k\right\},\label{eq:trun}
\end{equation}
where $I\left\{Y(\bs_i) \in G_k\right\}$ is an indicator function ensuring that $Y(\bs_i)$ belongs to a particular group, denoted $G_k$.  That is, we could model each household's income using a truncated normal distribution,
generating synthetic households that belong to the correct income brackets and preserving the proportions observed in the real population.  While such a model would reduce data privacy --- i.e., we must be willing to disclose a household's true income bracket --- data stewards \emph{know} this risk beforehand and can take appropriate measures.

While our work here was focused on scenarios with a single population and Gaussian outcomes, the framework we have presented can easily be extended to a multivariate framework and/or for use in generalized linear mixed models.  For instance, the value of a residence in San Francisco is likely a function of the location ($\bs_i$), number of bedrooms ($k$), the square footage ($\text{SqFt}_{1k}$), and the age of the property (in years; $\text{Age}_{2k}$).
To model the age of the property using differential smoothing, we could let
\begin{equation}
\text{Age}_{2k}(\bs_i)\given \gamma_0, \bw_{age} \sim Pois\left(\exp\left[\gamma_0 + w_{age}(\bs_i)\right]\right),
\end{equation}
where $\gamma_0$ is an intercept term and $w_{age}(\bs)$ is a differentially smoothed spatial process.
Then, to model the property's value, we could let
\begin{equation}
Y_k(\bs_i)\given \btheta_k \sim N\left(\beta_{0k} + \text{SqFt}_{1k}(\bs_i)\beta_{1k} + \text{Age}_{2k}(\bs_i)\beta_{2k}+ w_k(\bs_i), \tau_k^2\right),\;k=0,\ldots,K
\end{equation}
where $\btheta_k = \left(\bbeta_k,\bw_k,\tau_k^2\right)'$ and $\bw(\bs) = \left(w_0(\bs),\ldots,w_K(\bs)\right)'$ is a {differentially smoothed} multivariate spatial process.  In this model, predictions for an outlying observation at location $\bs_i$ with $k$ bedrooms would be based on its group-specific regression model, as well as a function of the observations near $\bs_i$ with a different number of bedrooms.  For instance, in a region comprised primarily of small condominiums, the spatial surfaces for studio (no bedroom) and one-bedroom units could help inform the surface for rarer two-bedroom units.

{We conclude by acknowledging that this work is just a first step toward achieving reduced disclosure risk.  One drawback of the restricted model used here is that it treats the idea of being a spatial outlier as a binary decision.  In our future work, we aim to devise an approach which defines $a_i$ continuously over the range $[0,1]$. One option would be to define
\begin{equation*}
a_i(\phi) = 1-\exp\left(-\phi\; \min_{j\ne i}||\bs_i-\bs_j||\right)
\end{equation*}
and $\gamma\left(\sig^2,\tau^2\right) =\sig^2\slash\tau^2 - 1$ as explicit functions of the parameters $\phi$, $\sig^2$, and $\tau^2$, and account for these definitions in our MCMC sampler.  While this is conceptually straightforward, it is unclear how such a framework would affect the convergence of our model parameters, much less whether these particular definitions are optimal.
}

\section*{Acknowledgements}
This research was partially supported by the U.S. National Science Foundation (NSF) and the U.S. Census Bureau under NSF grant SES-1132031, funded through the NSF-Census Research Network (NCRN) program.

\bibliographystyle{jasa}
\bibliography{mpp}

\end{document}